\documentclass[twocolumn,prl,aps,showkeys]{revtex4-2}
\usepackage{amsmath}
\usepackage{amsbsy}
\usepackage{latexsym,epsfig,graphicx}
\usepackage{dcolumn}
\usepackage{graphicx}
\usepackage{subfigure}
\usepackage{comment}
\usepackage{color}
\usepackage{bm}
\usepackage{mathrsfs}
\usepackage{amsfonts}
\usepackage{color}
\usepackage{amssymb}
\usepackage{xspace}
\usepackage{epstopdf}
\graphicspath{{figures/}}
\usepackage{tabularx}
\usepackage{longtable}
\usepackage{amsbsy}
\usepackage{float}
\usepackage{latexsym,epsfig,graphicx}
\usepackage{dcolumn}
\usepackage{graphicx}
\usepackage{subfigure}
\usepackage{comment}
\usepackage{color}
\usepackage{bm}
\usepackage{mathrsfs}
\usepackage{amsfonts}
\usepackage{amsmath}
\usepackage{color}
\usepackage{amssymb}
\usepackage{xspace}
\usepackage{epstopdf}
\usepackage{tabularx}
\usepackage{longtable}
\usepackage[colorlinks=true, letterpaper=true, pdfstartview=FitV, linkcolor=blue, citecolor=blue, urlcolor=blue]{hyperref}
\usepackage[normalem]{ulem}

\begin{document}
\title{Multiple mobility rings in non-Hermitian Su-Schrieffer-Heeger chain with quasiperiodic potentials}
\author{Guan-Qiang Li}
\thanks{Email: liguanqiang@sust.edu.cn}
\author{Zhi-Yu Lin, You-Jiao Dong, Ya-Feng Xue, Chun-Yang Ren}
\author{Ping Peng}
\affiliation{School of Physics and Information Science, Shaanxi University of Science and Technology, Xi'an 710021, China}
\affiliation{Institute of Theoretical Physics, Shaanxi University of Science and Technology, Xi'an 710021, China}
\affiliation{SUST-SpinQ Joint Laboratory for Quantum Computation, Shaanxi University of Science and Technology, Xi'an 710021, China}

\begin{abstract}
The localization property of a non-Hermitian Su-Schrieffer-Heeger (SSH) chain with quasi-periodic on-site potential is investigated. In contrast to the preceding investigations, the quantum phase transition between localized state and extended one is achieved by adjusting the strength of intracellular or intercellular hopping. The energy spectra and  eigenstate distributions of the system's Hamiltonian near the boundary of the phase transition exhibit different behaviors when the Hermiticity, non-Hermiticity and mosaic modulation of the quasi-periodic potential are considered, respectively. The existence of the mobility ring is revealed in the non-Hermitian SSH chain by studying of the critical behaviors near the boundary. More interestingly, the multiple mobility rings emerge when the period number of the mosaic modulation is increased. The result is helpful for the investigation of the localization-delocalization transition in the SSH-type system under the combined action of the non-Hermiticity and quasi-periodicity.
\end{abstract}


\maketitle

\section*{I. Introduction}
The discovery of Anderson localization has generated significant interest in the investigation of quantum states in disordered systems \cite{Anderson1958Absence, Abrahams1979Scaling, DasSarma1988, DasSarma1990, Lahini2008Anderson}. For one-dimensional (1D) or two-dimensional systems, even infinitesimally weak disorders can drive the whole spectra to complete localizations, while for the three-dimensional disordered systems it is found that the localized and extended states can coexist \cite{Evers2008Anderson, DasSarma2010, DasSarma2011}. The quantum phase transition between the localized state and extended one can be realized by adjusting the disorder intensity. The phase transition boundary in the energy spectrum is referred to as the mobility edge (ME) \cite{Li2015Many, Nag2017Many, Kohlert2019Observation, An2021Interactions}. After the discovery of the quasi-crystals \cite{Ds1984Metallic, Dl1984Quasicrystals}, the quasi-periodic systems have attracted much attention \cite{HH2016Fermipolaron, Wang2020Localization, Jagannathan2021The, Huang2021Fundamental}. The MEs exist even in low-dimensional quasi-periodic systems \cite{Aubry1980Analyticity, Soukoulis1982Localization, Kohmoto1983Metal, Sokoloff1985Unusual}. A typical example is the study of the Aubry-Andr\'{e} (AA) model, for which the quantum phase transition from the extended state to the localized one occurs at the critical strength of the quasi-periodic potential~\cite{Soukoulis1982Localization, Kohmoto1983Metal, Sokoloff1985Unusual}. Due to the special self-dual symmetry, the MEs can be solved analytically.

With the experimental realization and observation of localization phenomena in disordered systems \cite{Roati2008Anderson, Modugno2010Anderson}, a variety of novel localization-delocalization transitions have been gradually uncovered in the  generalized AA model, including the reentrant localization \cite{Roy2021Reentrant} and the many-body localization \cite{Iyer2013Many, Larcher2009Effects}. Certain systems endowed with special symmetry, such as parity-time (PT) symmetry, allow the exact determination of their MEs, which usually was a challenge for investigating the quantum phase transitions in the quasi-periodic and disordered systems \cite{SChen2020, Wang2020One, Tang2021Mobility, Zhang2022Lyapunov, Zhou2023Exact, Wang2024Non}. For the 1D systems, the MEs can alternatively be identified by calculating transmission coefficients, while the systems with higher-dimension require the utilization of the finite-size scaling theory by analyzing the change of the localization length with the system's size \cite{S.Bera2016DirtyWeyl}. Recently, the discovery of the mobility ring (MR) in the non-Hermitian AA models has overcome the limitation of the traditional MEs by precisely distinguishing localization-delocalization transitions \cite{Li2024Ring}. The advancement provides a more comprehensive framework for characterizing the phase transitions in non-Hermitian quantum systems.

Meanwhile, the topological phase transition has become one of the hot topics in frontiers of physics in recent years \cite{Kosterlitz2018Ordering, Thouless1982Quantized, Haldane1988Model, Kane2005Quantum, Bernevig2006Quantum,Qi2011Topological,CPMoca2022}. With the investigation of the topological Anderson insulators, it was believed that the disorder or quasi-periodic potential can trigger the transition from topological trivial phase to the nontrivial one \cite{Li2009Topological}. The generalized AA models with commensurate hopping can produce zero-energy topological states associated with Majorana fermions \cite{Alicea2012New, Gangadharaiah2011Majorana}. The incommensurate modulation affects the fate of the topological states \cite{Cestari2016Fate}. The Aubry-Andr\'{e}-Harper models with tunable phase shifts can exhibit the topologically nontrivial band structures and the symmetry-protected boundary states \cite{Liu2015Localization}. The energy-dependent metal-insulator transitions with a novel localized state can be realized in a 1D $p$-wave superconductor subject to slowly varying incommensurate potentials \cite{Liu2017Fate}.

The topological phase transition is also intensively studied in the Su-Schrieffer-Heeger (SSH) model with intracellular and intercellular hoppings \cite{asboth2016short, bergh2021Exceptional}. Current discussions about the effect of the quasi-periodicity on the phase transition in the SSH model focus on tuning the strengths of the on-site potential and the quasi-periodic hoppings between different sites \cite{Miranda2024Mechanical, Sircar2024Topological, Liu2022Anomalous}. Meanwhile, more and more exotic topological properties have been revealed in non-Hermitian SSH systems~\cite{K.Sun2025, Lima2025}. In the present article, we investigate the localization properties of the quantum states in the non-Hermitian SSH model with the quasi-periodic on-site potential by adding mosaic modulation. When the period number of the mosaic modulation is larger than two, the  multiple mobility rings (MMRs) are obtained analytically. The number of the rings increases with increasing the period number of the modulation. The behaviors of the localization-delocalization transition under specific conditions are discussed by adjusting the strength of the intercellular or intracellular hopping. Furthermore, we examine the spatial distribution of the modulus of the eigenstates for analyzing the effect of the non-Hermitian term on the localization properties of the model. For the non-Hermitian situation, the MR can accurately distinguish the localized state and the extended one near the phase boundaries~\cite{Li2024Ring}. The MMR presenting as ring structure in the complex energy plane can be thought of as the generalization of the ME for the non-Hermitian quasi-periodic potentials.

\section*{II. Basic model of non-Hermitian SSH chain with quasiperiodic potential}
\begin{figure}[h]
\includegraphics[width=0.49\textwidth]{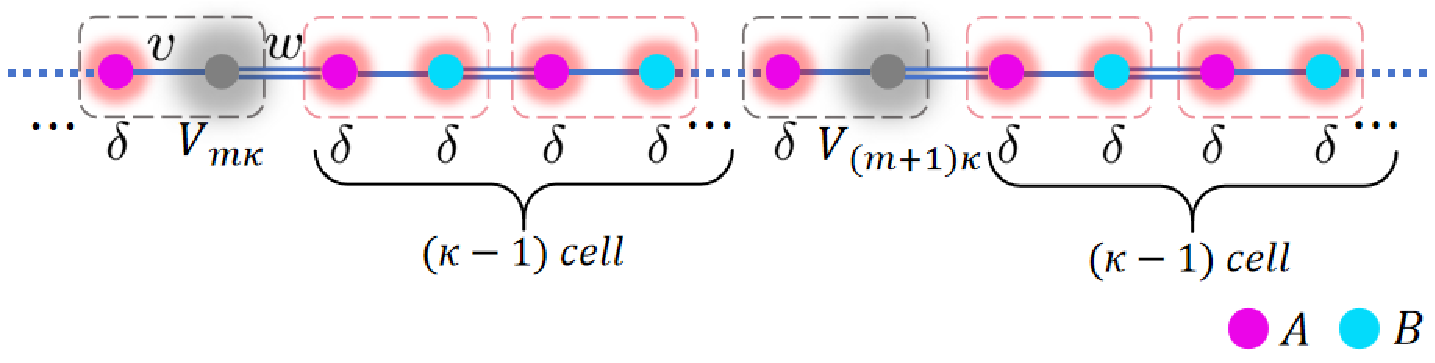}
\caption{Schematic diagram of the SSH model for the quasi-periodic potential $V_{j}$ with the mosaic modulation. Two sites per unit cell are represented by $A$ and $B$, $v$ represents the strength of the intracellular hopping, while $w$ denotes the strength of the intercellular hopping. }\label{figure1}
\end{figure}
We consider the SSH model with quasi-periodic on-site potential, which is shown in Fig.~\ref{figure1}. The Hamiltonian of the model is given as follows
\begin{gather} 
H=\sum_{j=1}^{L/2}(vc_{j,A}^{\dagger}c_{j,B}+vc_{j,B}^{\dagger}c_{j,A}+V_{j,A}c_{j,A}^{\dagger}c_{j,A}+V_{j,B}c_{j,B}^{\dagger}c_{j,B})
\notag\\
+\sum_{j=1}^{L/2-1}w(c_{j+1,A}^{\dagger}c_{j,B}+c_{j,B}^{\dagger}c_{j+1,A}),
\label{Hamiltonian2}
\end{gather}
with $V_{j,A}=\delta$ and
\begin{equation}
V_{j,B}=\left\{
\begin{array}{lll}
2\lambda \cos(2\pi \alpha j+\theta+ih)&,j=m\kappa\\
\delta  &,  other\\
\end{array}
\right.
\end{equation}
Here, $c_{j, A}^{\dagger}$($c_{j, A}$) and $c_{j, B}^{\dagger}$($c_{j, B}$) represent the creation (annihilation) operators on the sites $A$ and $B$ in the $j$th cell, respectively. The hopping strength between two adjacent sites in the same cell is denoted by $v$ and the strength between two adjacent cells by $w$. The model consists of $L/2$ cells and the number of the sites is $L$. $V_{j,A/B}$ represents the on-site potential, in which the quasi-periodicity, non-Hermicity and mosaic effects could be included together. The on-site potential is quasi-periodic once $\alpha$ is taken as an irrational number. $\lambda$ represents the intensity of the  potential. $\theta+ih$ denotes a complex phase offset, which makes the quasi-periodic potential non-Hermitian. $\kappa$ is a positive integer that determines the period of the mosaic modulation superposing on the quasi-periodic potential. The other part of the on-site potential, represented by $\delta$, is set to be a real or complex constant. In the following numerical calculations, the site number is chosen as $L=2F_{15}$, in which $F_{15}=610$ is the $15$th Fibonacci number. The intensity of the quasi-periodic potential is set to $\lambda=0.5$ and the irrational number $\alpha=(\sqrt{5}-1)/2$.

The Lyapunov exponent (LE) can be determined by the transfer matrix method, which rigorously characterizes the localization properties of the eigenstates~\cite{Wang2020One, Li2024Ring}. We obtain the analytical expressions of the corresponding MRs in Eqs.~(\ref{E11}) and (\ref{E18}) for $\kappa=1$ and $\kappa=2$, respectively. The process is given as follows. The LE is defined as
\begin{gather}
\gamma_{h}(E)=\lim_{N \to \infty} \frac{1}{2\pi N} \int \ln_{}{\left \|\mathcal{T}_{N}\left ( \theta  \right )   \right \| } \mathrm{d}\theta,
\end{gather}
where $\mathcal{T}_{N} =\prod_{m=1}^N T_{m} = T_{N} T_{N-1} \cdots T_{2}T_{1}$, $T_{m}$ is the transfer matrix of a quasicell with one cell interacted by the mosaic modulation. Here, $N\equiv L/(2\kappa)$ denotes the number of the quasicells, $m=1,2,\ldots,N$ as the quasicell index, and $\|\cdot\|$ represents the norm of matrix. The transfer matrix can be given as
\begin{gather}
T_{m}=T_{B} T_{A} T_{\kappa},
\end{gather}
with
\begin{equation}
	T_{A}=\left(
	\begin{array}{lll}
		\frac{E-\delta}{v}&~~~-\frac{w}{v}\\
		\\
		~~1&~~~~~~0\\
	\end{array}
			\right),
	T_{B}=\left(
	\begin{array}{lll}
		\frac{E-V_{m\kappa,B}}{w}&~~~-\frac{v}{w}\\
		\\
		~~~1&~~~~~~0\\
	\end{array}
	\right),\label{}
\end{equation}
and
\begin{equation}
	T_{\kappa}=\left(
	\begin{array}{lll}
		\frac{(E-\delta)^{2}-v^{2}}{vw}&~~~-\frac{E-\delta}{v}\\
		\\
		~~~\frac{E-\delta}{v}&~~~-\frac{w}{v}\\
	\end{array}
	\right)^{\kappa-1}.\label{}
\end{equation}
By letting $h\to \infty$, we can obtain for the mosaic modulation with $\kappa=1$
\begin{equation}
T_{m,h\to \infty}=e^{-i(2\pi\alpha j+\theta)}e^h \left(
\begin{array}{lll}
\frac{-\lambda (E-\delta)}{vw}&~~~\frac{\lambda}{v}\\
\\
~~~~~0&~~~0\\
\end{array}
\right)+o(1).\label{E9}
\end{equation}
Thus, we have
\begin{gather}
\gamma_{h\to \infty}(E)=\ln_{}{\left | \frac{\lambda (E-\delta)}{vw} \right |  } + h + o(1).\label{E10}
\end{gather}

According to the Avila's global theory, $\gamma_{h}(E)$ is a convex, piecewise-linear function of $h$ with integer slopes~\cite{Avila2015Global}. For the present model, the possible value of the slopes is either $0$ or $1$. Furthermore, $\gamma_{h}(E)$ is an affine function in a neighborhood of $h=0$ for any of $E$ and it means $\gamma_{h=0}(E)=0$. The LE for all real $h$ can only be expressed as follows
\begin{equation}
\gamma_{h}(E)=\mathrm{Max} \left\lbrace \ln_{}{\left|\frac{\lambda (E-\delta)}{vw}\right|}+|h|, 0 \right\rbrace.
\end{equation}
$\gamma_{h}(E)>0$ corresponds to the localized state and $\gamma_{h}(E)=0$ to the extended one in the thermodynamic limit $N\rightarrow\infty$. So we can obtain the critical condition for the boundaries by setting $\gamma_{h}(E)=0$, i.e., $\lambda (E-\delta) e^{ h}=vw$. If we set $E=E_{R}+iE_{I}$, where $E_{R}$ and $E_{I}$ are the real and imaginary parts of the eigenvalues, the general expression of the MR in the complex plane is obtained as
\begin{gather}
x^2+y^2=(\frac{vw}{\lambda e^{ h}})^2,\label{E11}
\end{gather}
where $x=E_{R}-\delta_{R}$ and $y=E_{I}-\delta_{I}$. $\delta_{R}$ and $\delta_{I}$ are the real and imaginary parts of $\delta$. The trajectory of the MR in complex plane is a circle centered at
$(\delta_R,\delta_I)$ with radius $vw/(\lambda e^{h})$.

For the mosaic modulation with $\kappa=2$, we have
\begin{equation}
	T_{m,h\to \infty}=e^{-i(2\pi\alpha j+\theta)}e^h \left(
	\begin{array}{lll}
		-\frac{\lambda}{w}f&~~~0\\
		\\
		~~~0&~~~0\\
	\end{array}
	\right)+o(1),
\end{equation}
with
\begin{gather}
f=\frac{(E-\delta)^3}{v^2w}-\frac{E-\delta}{w}-\frac{(E-\delta) w}{v^2}.\nonumber
\end{gather}
Thus, we have
\begin{gather}
\gamma_{h\to \infty}(E)=\ln_{}{\left | \frac{\lambda}{w}f \right |  } + h + o(1).\label{E17}
\end{gather}
Finally, one can obtain the general expression of the MMR in the complex plane
\begin{gather}
	\frac{(x^2+y^2)^2}{v^4 w^2}- \frac{2 (x^2-y^2)}{v^2w^2}+\frac{1}{w^2}
	 -\frac{2 (x^2+y^2)-w^2}{v^4}\notag \\
	 +\frac{2}{v^2}
	 = \frac{1}{ x^2+y^2} (\frac{w}{e^{\lambda h}})^2.\label{E18}
\end{gather}
Eq.~{(\ref{E18})} exhibits continuous symmetries of $SO(2)$ rotation, mirror reflection, and simultaneous scaling of $(v, w)$. In the $(x,y)$-plane, the MMR appears as concentric circular rings that can merge or vanish as the parameters $(v, w)$ are varied. The factor $e^{-2\lambda h}$ causes the radii to contract exponentially. The analytical expression of the MMR for $\kappa=2$ is the most important result in the present study.

\section*{\uppercase\expandafter{\romannumeral3}. Phase transition between localized and extended states }
\begin{figure}[htbp]
\includegraphics[scale=0.44,trim=45 40 0 0,clip]{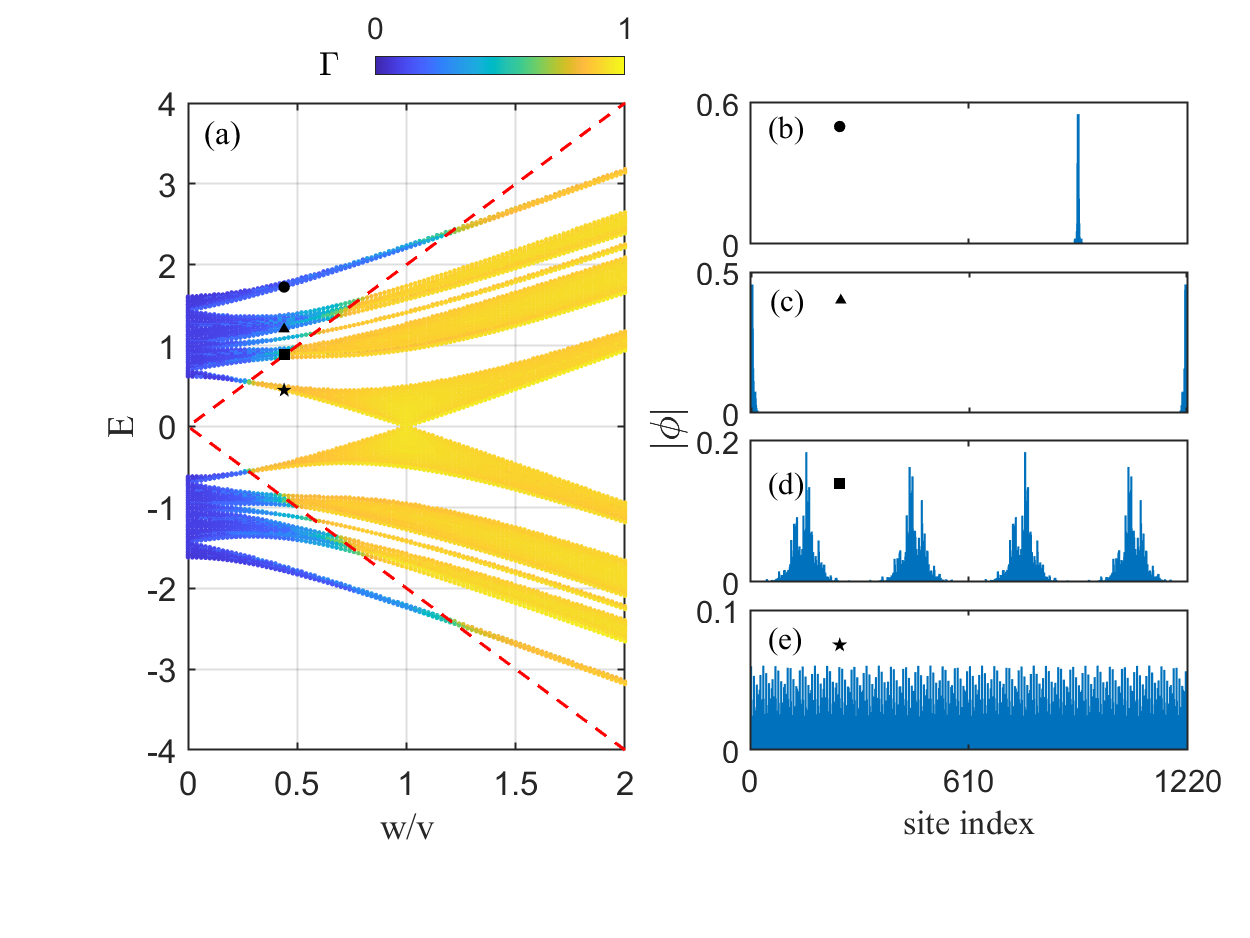}
\caption{ (a) Change of the fractal dimension $\Gamma$ with the energy $E$ and the strength ratio $w/v$. The two red dashed lines mark the exact MEs. (b)-(e) The spatial distributions of the four typical eigenstates corresponding to the eigenvalues marked by black circle, triangle, square and star in (a) when $w/v=0.44$. The other parameters are set to $\kappa=1$, $h=0$ and $\delta=0$. }
\label{figure2}
\end{figure}

We first consider the effect of intracellular and intercellular hoppings on the properties of localization for the SSH model with Hermitian quasi-periodic potential $(h=0)$. The fractal dimension $\Gamma$, which is usually used to characterize the properties of the localization~\cite{Wang2024Non, Wang2020One, Li2024Ring}, is defined as
\begin{gather}
\Gamma= - \lim_{L \to \infty} \frac{\ln_{}{(\sum_{j}^{}\sum_{\sigma={A,B}}|\phi_{j,\sigma}|^4)}}{\ln_{}{L}},\label{Gamma}
\end{gather}
where $\phi_{j,\sigma}$ denotes the probability amplitude of the eigenstate at the site $\sigma$ for the $j$-th cell. For the mosaic modulation with $\kappa=1$, the distribution diagram of the fractal dimension $\Gamma$ with the energy $E$ and the strength ratio of the hoppings $w/v$ is shown in Fig.~\ref{figure2}(a). The distribution of the fractal dimension is symmetric about $E=0$. The energy gap closes at $w/v=1$. Importantly, the diagram can be divided into two distinct regions by identifying the magnitude of the fractal dimension. One represents the localized states with $\Gamma\rightarrow0$ (blue part in the diagram), while the other corresponds to the extended states with $\Gamma\rightarrow1$ (yellow part in the diagram). The phase boundary of the localization-delocalization transition can be verified by the exact ME, which is denoted by the red dashed lines and given analytically by Eq.~(\ref{E11}). It is found that the energy on the lines changes linearly with $w/v$. With increasing $w/v$, the system may undergo quantum phase transition from the localized state to the extended one by suitably choosing the specific energy. Fig.~\ref{figure2}(b)-(e) give the spatial distributions of the eigenstates of the model, corresponding to four marks with the energies $E=1.73,1.21,0.89,0.45$ and $w/v=0.44$ shown in Fig.~\ref{figure2}(a). Fig.~\ref{figure2}(b) and (c) illustrate the features of the localized states with one peak and two peaks. Since the eigenstates are obtained under the periodic boundary condition, the two peaks in Fig.~\ref{figure2}(c) are actually equivalent to each other. Fig.~\ref{figure2}(e) illustrates the typical feature of an extended state. Fig.~\ref{figure2}(d) displays the eigenstate at the boundary between localized and extended phases. There are four isolated peaks with larger widths for the eigenstate, which is the crossover state between the localized state and the extended one. Fig.~\ref{figure2}(b)-(e) give us the knowledge about the evolution process of the eigenstate across the phase boundary.
\begin{figure}[htbp]
\includegraphics[scale=0.44,trim=45 40 0 0,clip]{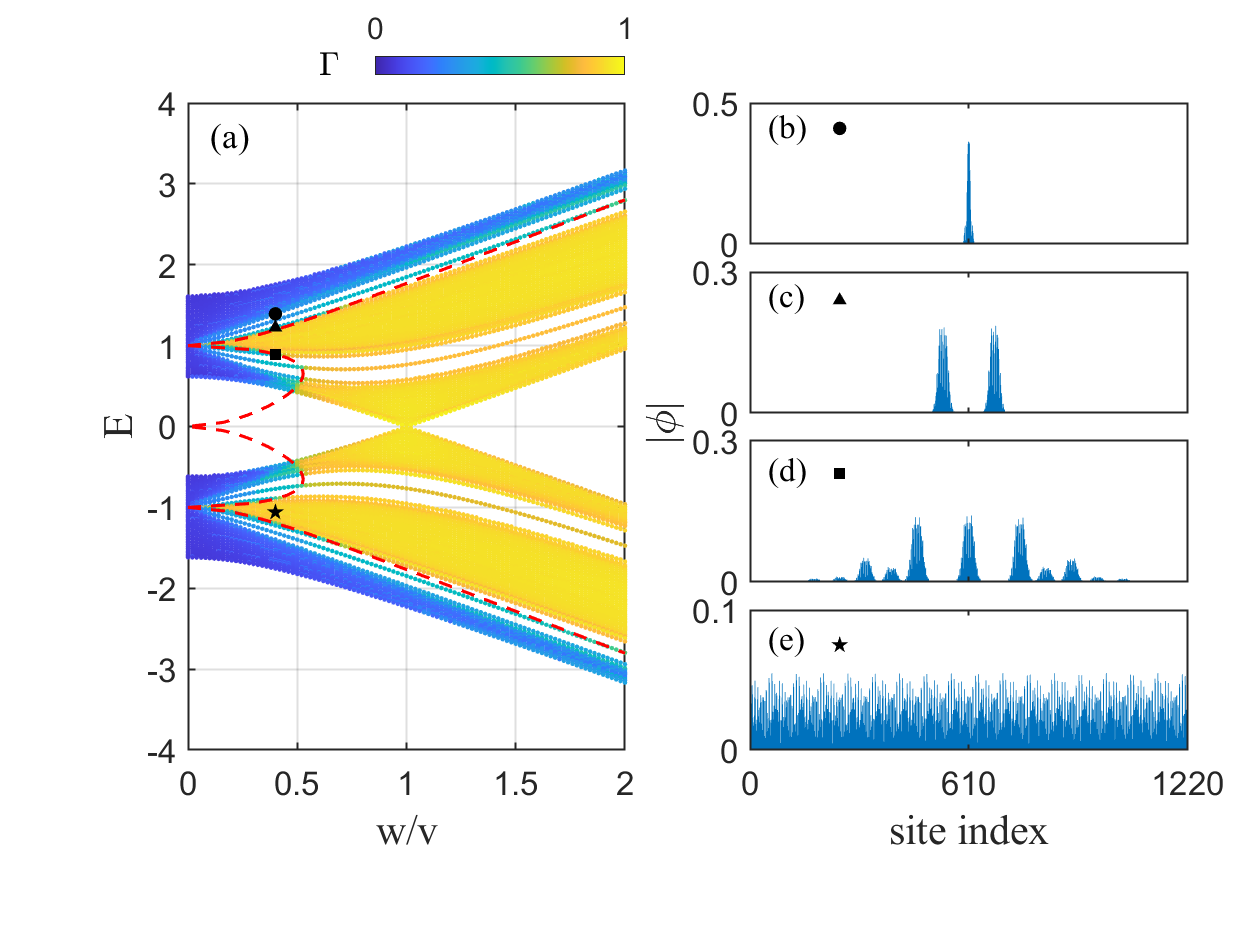}
\caption{(a) Change of the fractal dimension $\Gamma$ with the energy $E$ and the strength ratio $w/v$, the red dashed lines denote the exact MEs. (b)-(e) The spatial distributions of the four typical eigenstates corresponding to the eigenvalues marked by black circle, triangle, square and star in Fig.~\ref{figure3}(a) when $w/v=0.4$. The other parameters are set to $\kappa=2$, $h=0$ and $\delta=0$. }
\label{figure3}
\end{figure}

For the mosaic modulation with $\kappa=2$, the distribution diagram of the fractal dimension $\Gamma$ with the energy $E$ and the strength ratio $w/v$ is illustrated in Fig.~\ref{figure3}(a). The diagram is also divided into two distinct regions by the fractal dimension, corresponding to the localized states with $\Gamma\rightarrow0$ (blue part in the diagram) and the extended states with $\Gamma\rightarrow1$ (yellow part in the diagram). Compared to the case of the mosaic modulation with $\kappa=1$, the system has more complex phase boundary for the case of $\kappa=2$. The phase boundaries of the localization-delocalization transition are denoted by the red dashed curves in Fig.~\ref{figure3}(a) and given analytically by Eq.~(\ref{E18}). Obviously, the change of the energy on the boundary with $w/v$ demonstrates a non-monotonic behavior and the energy has no one-to-one correspondence relationship with $w/v$ for about $w/v\leq0.53$. Fig.~\ref{figure3}(b)-(e) give the spatial distributions of the eigenstates of the Hamiltonian, corresponding to four marks with the energies $E=1.39,1.22,0.89,-1.06$ and $w/v=0.4$ shown in Fig.~\ref{figure3}(a). Fig.~\ref{figure3}(b) gives the typical distribution of the localized state with $\Gamma\rightarrow0$ while Fig.~\ref{figure3}(e) gives the typical distribution of the extended state with $\Gamma\rightarrow1$. Although Fig.~\ref{figure3}(c) and (d) have different distributions, they are the crossover states between the localized state and the extended one since the parameters are chosen on the phase boundary.

\begin{figure}[htbp]
\includegraphics[scale=0.44,trim=45 95 0 0,clip]{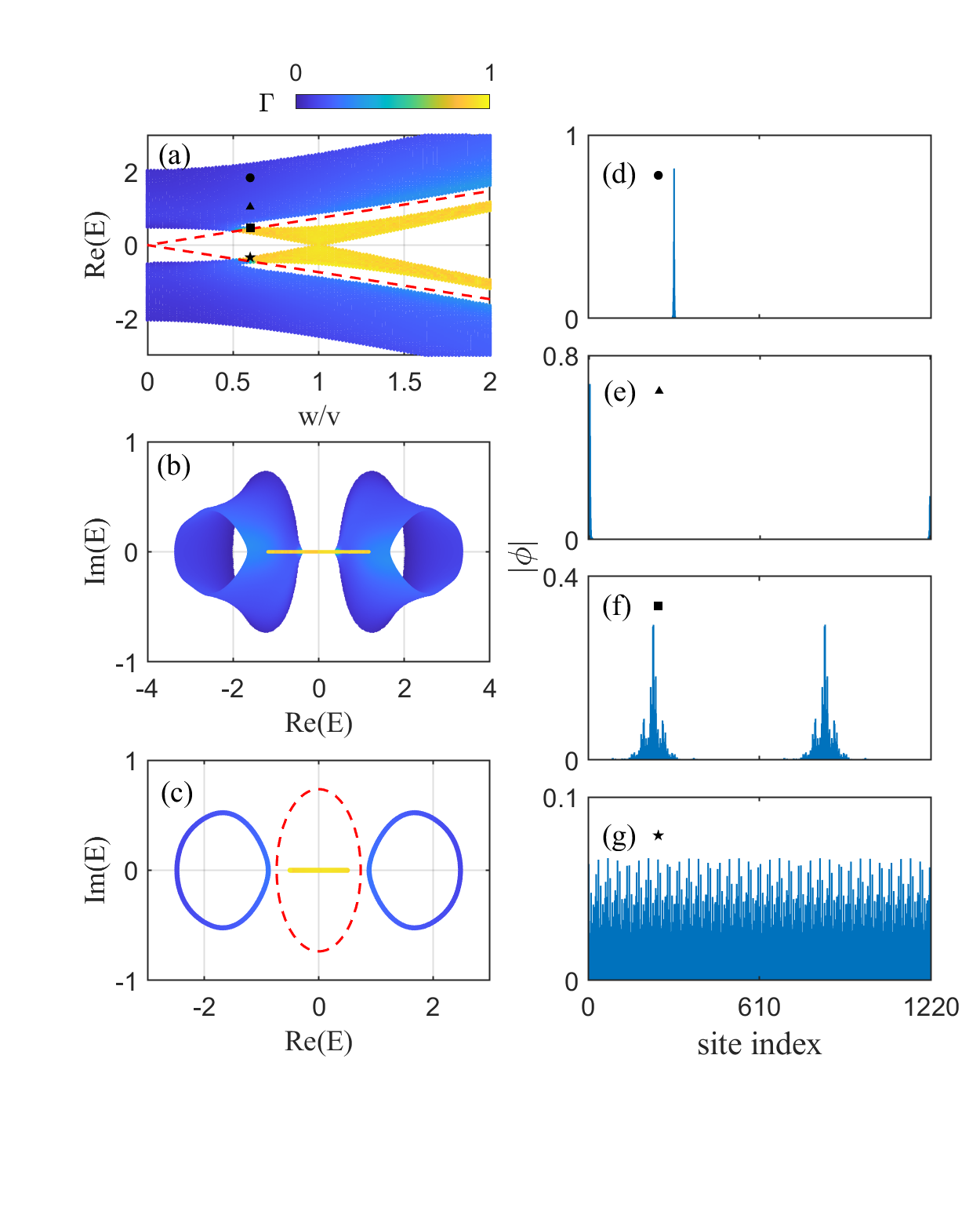}
\caption{(a) Change of the fractal dimension $\Gamma$ with the energy $Re(E)$ and the strength ratio $w/v$. The red dashed lines are the exact MEs. (b) The fractal dimension $\Gamma$ versus $Re(E)$ and $Im(E)$ over the range of $w/v\in[0,2]$. (c) The distribution of the fractal dimension on the complex plane constructed by the real and imaginary parts of the energy when $w/v=1.0$. The red dashed line denotes the MR obtained from Eq.(\ref{E11}). (d)-(g) The spatial distributions of the four typical eigenstates corresponding to the eigenvalues marked by black circle, triangle, square and star in Fig.~\ref{figure4}(a) when $w/v=0.6$. The other parameters are set to $\kappa=1$, $h=1$ and $\delta=0$.}
\label{figure4}
\end{figure}

We now investigate the localization behaviors of the model with non-Hermitian quasi-periodic potential $(h\neq0)$. The result can be obtained by solving Eq.~(\ref{Hamiltonian2}) when the parameters are chosen the same as in Fig.~\ref{figure2}(a) except for $h=1$. The non-Hermitian model has complex energy spectrum. For the mosaic modulation with $\kappa=1$, the distribution diagram of the fractal dimension $\Gamma$ with the real part of the energy $\mathrm{Re}(E)$ and the strength ratio $w/v$ is shown in Fig.~\ref{figure4}(a). The distribution of the fractal dimension is symmetric about $\mathrm{Re}(E)=0$. Compared to the Hermitian case, the model in this situation exhibits distinct localization characteristics. For the region of $\mathrm{Re}(E)>0$, the magnitude of the fractal dimension equals to zero when $w/v$ is smaller than a critical value of $\sim0.54$. With increasing $w/v$, the regions for the localized states with $\Gamma\rightarrow0$ (blue part in the diagram) and the extended states with $\Gamma\rightarrow1$ (yellow part in the diagram) are divided by a gap, which is the boundary of the localized-extended phase transition. The exact ME is given analytically by Eq.~(\ref{E11}) and  illustrated by the red dashed line across the gap in Fig.~\ref{figure4}(a). The change of the fractal dimension with the real part $\mathrm{Re}(E)$ and the imaginary part $\mathrm{Im}(E)$ of the energy over the range of $w/v\in[0, 2]$ is given in Fig.~\ref{figure4}(b). Besides the two blue toruses symmetrically distributed at both sides of $\mathrm{Re}(E)=0$, there exists a yellow line segment with $\mathrm{Im}(E)=0$ representing for the extended states. For given $w/v=1.0$, the two toruses degrade into two rings on the plane of the energy, as shown in Fig.~\ref{figure4}(c). The eigenstates on the rings are localized and on the yellow line segment with $\mathrm{Im}(E)=0$ are extended. The boundary between the localized states and the extended ones can be determined from Eq.~(\ref{E11}), which is demonstrated by the red dashed circle with radius $2.0/e$ for the specific parameters in Fig.~\ref{figure4}(c). Fig.~\ref{figure4}(d)-(g) give the spatial distributions of the eigenstates, corresponding to four marks with the energies $E=1.83,  1.04,  0.47, -0.33$ and $w/v=0.6$ shown in Fig.~\ref{figure4}(a). Fig.~\ref{figure4}(d) and (e) give the typical distributions of the localized states with $\Gamma\rightarrow0$ while Fig.~\ref{figure4}(g) gives the typical distribution of the extended state with $\Gamma\rightarrow1$. Since the eigenstate is obtained under the periodic boundary condition, the two peaks in Fig. ~\ref{figure4}(e) are actually equivalent to one, which is similar to the case in Fig.~\ref{figure2}(c). The distribution in Fig.~\ref{figure4}(f) is the crossover state between the localized state and the extended one.

Let us further explain the origin and significance of Eqs.~(\ref{E11}) and (\ref{E18}). For the Hermitian quasi-periodic potential, the phase boundary between the localized and extended states is given analytically by the ME. Two examples are demonstrated by the red dashed lines in Fig.~\ref{figure2}(a) and Fig.~\ref{figure3}(a). For the non-Hermitian quasi-periodic potential, the boundary between the localized and extended states is determined analytically by the so-called MR, which is represented by the red dashed circle in Fig.~\ref{figure4}(c). The study of the ME in the low-dimensional quasi-periodic system has a long history~\cite{DasSarma1988, DasSarma1990, DasSarma2010, Li2015Many, Wang2020One, Zhou2023Exact}. The generalization of the ME to the situation of the non-Hermitian quasi-periodic system has become the hot topic in recent years~\cite{Li2024Ring, Wang2024Non, Chen2025Mobilityrings, wang2025exactmultiple}. It is found that the phase boundary between the localized and extended states in the non-Hermitian quasi-periodic systems can be fully described by the MR~\cite{Li2024Ring}. Our research will further extend the finding.
\begin{figure}[htbp]
\includegraphics[scale=0.44,trim=45 40 0 0,clip]{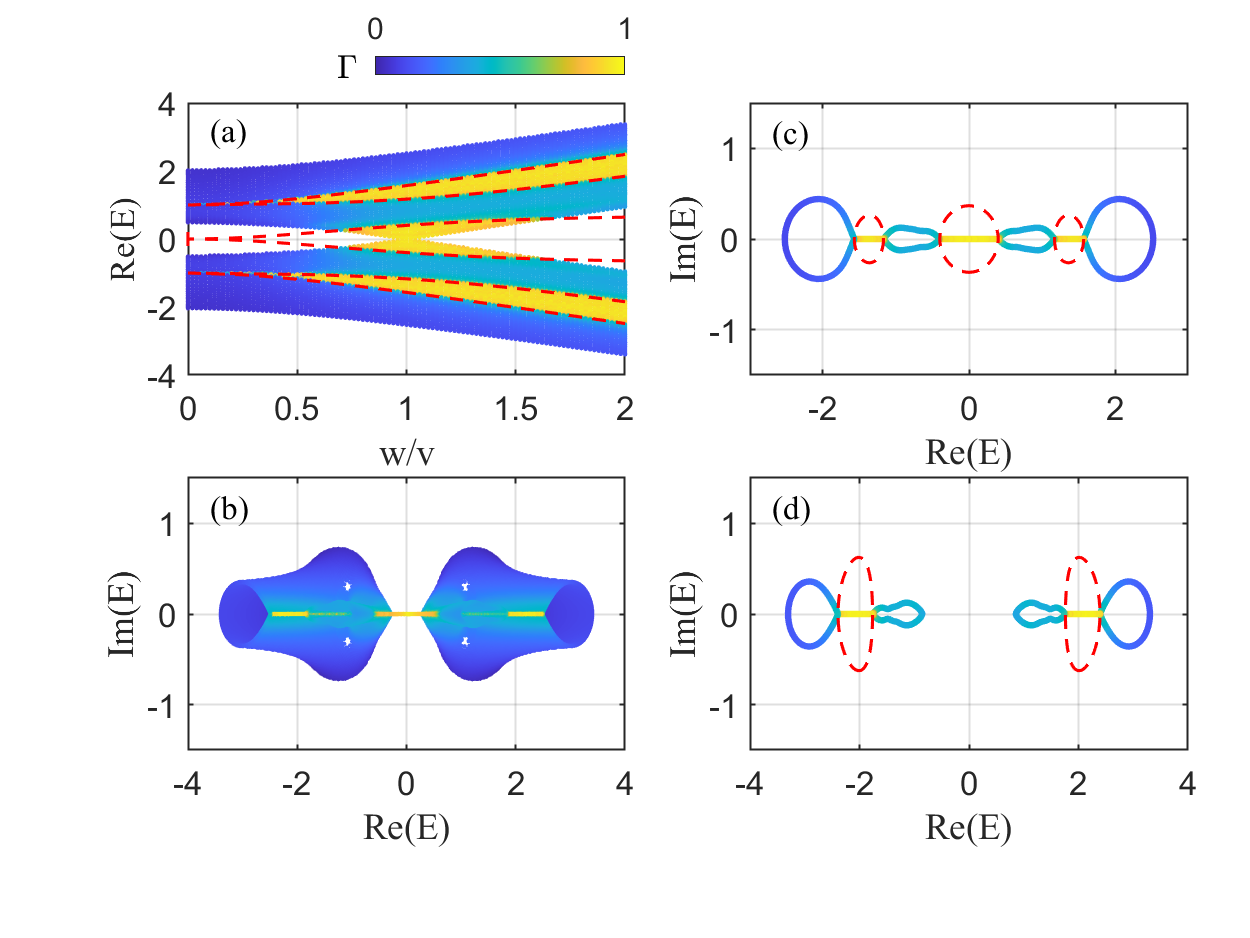}
\caption{(a) Change of the fractal dimension $\Gamma$ with the energy $Re(E)$ and the strength ratio $w/v$. The red dashed lines are the exact MEs. (b) The fractal dimension $\Gamma$ versus $Re(E)$ and $Im(E)$ over the range of $w/v\in[0,2]$. (c) and (d) The distribution of the fractal dimension on the complex plane constructed by the real and imaginary parts of the energy when $w/v=1$ and $w/v=1.9$. The red dashed lines are the MRs obtained from Eq.~(\ref{E18}). The other parameters are set to $\kappa=2$, $h=1$ and $\delta=0$. }
\label{figure5}
\end{figure}

\section*{\uppercase\expandafter{\romannumeral4}. MMR induced by the periodicity of the mosaic modulation}
For the non-Hermitian potential with $\kappa=2$, the change of the fractal dimension $\Gamma$ with the real part of energy $\mathrm{Re}(E)$ and the strength ratio $w/v$ is presented in Fig.~\ref{figure5}(a). The distribution of the fractal dimension is still symmetric about $\mathrm{Re}(E)=0$. The phase boundary of the localization-delocalization transition is determined analytically by Eq.~(\ref{E18}) and indicated by the red dashed curves in Fig.~\ref{figure5}(a). There are three pairs of phase boundaries. Different from what is observed for the Hermitian potential with $\kappa=2$ in Fig.~\ref{figure3}(a), the change of the real part of energy $\mathrm{Re}(E)$ versus $w/v$ shows a monotonic behavior on each phase boundary. For given $w/v$, the reentrant localization can be realized by changing the energy of the system~\cite{Roy2021Reentrant}. The distribution of the fractal dimension $\Gamma$ with the real part of energy $\mathrm{Re}(E)$ and imaginary part of energy $\mathrm{Im}(E)$ over the range of $w/v\in[0,2]$ is illustrated in Fig.~\ref{figure5}(b). The distribution of the fractal dimension is symmetric about $\mathrm{Re}(E)=0$ and $\mathrm{Im}(E)=0$. Combined analysis of Fig.~\ref{figure5}(a) and (b) indicates that the complex structure on the energy plane may evolve into multiple loops as $w/v$ is changed. The two cases for a direct visual comparison are presented by Fig.~\ref{figure5}(c) and (d) with $w/v=1$ and $w/v=1.9$. The boundary of the localized-extended phase transition is determined analytically by Eq.~(\ref{E18}) and indicated by the red dashed rings on the complex plane about the energy. These dashed rings are called as the MRs. There are three MRs in Fig.~\ref{figure5}(c) and two MRs in Fig.~\ref{figure5}(d), depending on the different choice of the strength ratio $w/v$. The MMR structure with the ring number larger than three lays the foundation for the investigation of the reentrant localization in the SSH model by adjusting the strength of the intercellular or intracellular coupling.
\begin{figure}[htbp]
\includegraphics[scale=0.44,trim=45 40 0 0,clip]{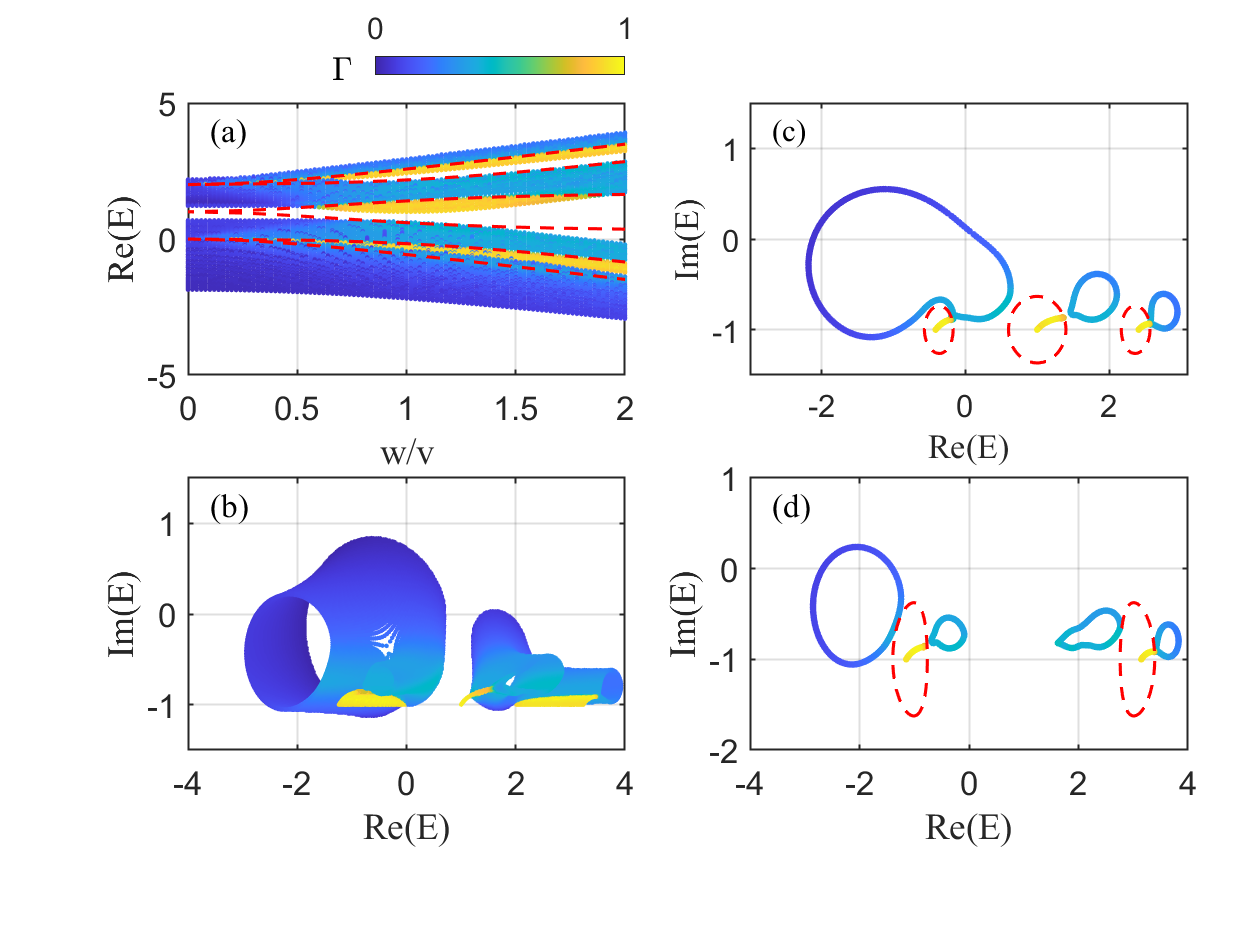}
\caption{(a) Change of the fractal dimension $\Gamma$ with the energy $Re(E)$ and the strength ratio $w/v$. The red dashed lines are the exact MEs. (b) The fractal dimension $\Gamma$ versus $Re(E)$ and $Im(E)$ over the range of $w/v\in[0,2]$. (c) and (d) The distribution of the fractal dimension on the complex plane constructed by the real and imaginary parts of the energy when $w/v=1$ and $w/v=1.9$, respectively. The red dashed lines are the MRs obtained from Eq.~(\ref{E18}). The other parameters are set to $\kappa=2$, $h=1$ and $\delta=1+i$. }
\label{figure6}
\end{figure}

Non-Hermicity of the system can also be induced by the complex on-site potential $\delta$. Compared with the non-Hermicity induced by $h\neq0$, the complex $\delta$ make the non-Hermitian system to be PT-symmetry broken. For the non-Hermitian mosaic potential with the same parameters as that in Fig.~\ref{figure5} except for $\delta=1+i$, the distribution of the fractal dimension $\Gamma$ with the real part of energy $Re(E)$ and the strength ratio $w/v$ is presented in Fig.~\ref{figure6}(a). The distribution does not keep symmetric about $\mathrm{Re}(E)=0$ and the band gap does not close when $w/v=1$ is satisfied. Not only that, due to the PT-symmetry breaking of the non-Hermitian mosaic potential, the analytically determined red dashed curves deviate from the actual phase boundary. Fig.~\ref{figure6}(b) shows the distribution of the fractal dimension as a function of the real and imaginary parts of the energy over the range of $w/v\in[0,2]$. The complex energy spectra become several irregular toruses and the distribution of the fractal dimension do not exhibit symmetry about $\mathrm{Re}(E)=0$ and $\mathrm{Im}(E)=0$. In Fig.~\ref{figure6}(c), the complex energy spectra degrade into three irregular loops and three line segments for $w/v=1$. But the spectra degrade into four irregular loops and two line segments for $w/v=1.9$ in Fig.~\ref{figure6}(d). The states with the energy on the loops are localized since $\Gamma\rightarrow0$ (blue part in the diagram), while the states with energy on the line segments are extended since $\Gamma\rightarrow1$ (yellow part in the diagram). Accordingly, there are three MRs in Fig.~\ref{figure6}(c) and two MRs in Fig.~\ref{figure6}(d), which are determined analytically by Eq.~(\ref{E18}) and represented by the red dashed rings. The extended state is located within the inner region of the MR, while the localized state is situated outside the MR. One can always obtain the MMR by adjusting the ratio $w/v$. For the cases with the non-Hermitian mosaic potentials, the MMRs correspond to the multiple boundaries of localized-extended phases in the complex energy spectra via the distribution of the fractal dimension. The MMR provides a powerful tool for analyzing and potentially controlling quantum state localization in non-Hermitian quasi-periodic systems. The non-Hermiticity of the on-site potential is the origin of the MR, the mosaic modulation leads to the multiplicity of the ring for suitably choosing strength ratio of the intercellular and intracellular couplings. It is worth emphasizing that the MMR with the ring number larger that three can be obtained by increasing the period number of the mosaic modulation in our system. Although it is difficult to obtain the analytical expressions for the MMR when $\kappa\geq3$, it is always possible to get the numerical solutions. 

\section*{\uppercase\expandafter{\romannumeral5}. Conclusion }
In this article, we investigate the localization behaviors of the non-Hermitian SSH chain with mosaic-modulated quasi-periodic potential. The system can exhibit rich behaviors of the localization-delocalization phase transition by adjusting the strengths of the non-Hermiticity and intercellular/intracellular hopping. The boundary of the quantum phase transition is verified by the exact ME in the Hermitian case, which exhibits increased complexity as increasing the periodicity of the mosaic modulation. For the non-Hermitian case, the existence of the MR is verified and the exact MR is determined analytically by the transfer matrix method. Especially, the MMR is obtained when the period number of the mosaic modulation is increased. In addition, the symmetry breaking induced by the non-Hermiticity of the potential is revealed in the phase diagram and energy spectra. These findings not only deepen our understanding of the phase transitions in quasiperiodic systems but also provide new insights for controlling the localization of quantum state, holding significant implications for the study of localization phenomena in non-Hermitian systems.

\section*{Acknowledgments}
We thank Zhi Li for his insightful discussions. The research is supported by NSF of China (Grant No. 11405100) and the Natural Science Basic Research Plan in Shaanxi Province (Grant Nos. 2019JM-332 and 2020JM-507).

\end{document}